\documentclass{article}
\usepackage{authblk}
\title{Master symmetries of non-linear systems}

\author{Nitin Serwa}

\affil{Department of Mathematics and Statistics, Abu Dhabi University, Al Ain, Abu Dhabi, UAE\\\texttt{nitin.serwa@adu.ac.ae}}

\usepackage{amssymb}
\usepackage{amssymb, amsthm, amsmath, graphicx, etoolbox, hyperref, amsxtra, amstext, tikz-cd, setspace, listings,mathrsfs}

\theoremstyle{definition}
\newtheorem{definition}{Definition}
\newtheorem{theorem}{Theorem}
\newtheorem{example}{Example}

\renewcommand{\setminus}{\smallsetminus}
\newcommand{\derf}[2]{\frac{\partial#1}{\partial#2}}
\newcommand{\ad}{\texttt{ad}}
\newcommand{\curly}[1]{\mathscr{#1}}
\newcommand{\cum}{{\textstyle \int}}
\newcommand{\Sl}{\mathfrak{sl}}
\newcommand{\CC}{\mathbb{C}}

\newcount\colveccount 
\newcommand*\colvec[1]{
        \global\colveccount#1
        \begin{pmatrix}
        \colvecnext
}
\def\colvecnext#1{
        #1
        \global\advance\colveccount-1
        \ifnum\colveccount>0
                \\
                \expandafter\colvecnext
        \else
                \end{pmatrix}
        \fi
}

\begin{document}
\newpage
\maketitle
\vspace{18mm} \setcounter{page}{1} \thispagestyle{empty}

\begin{abstract}
We explore new symmetries in two-component third order Burgers' type systems in (1+1)-dimension using Wang's $\mathscr{O}$-scheme. We also find a master symmetry for a (2+1)-dimensional Davey-Stewartson type system. These results shed light on the behavior of these equations and help us understand their integrability properties. Our approach offers a practical method for identifying symmetries, contributing to the study of integrable systems in mathematics and physics.
\newline \textbf{Keywords}: integrability, master symmetry, symmetry, homogeneous integrable nonlinear equation
\end{abstract}
\maketitle
\pagestyle{plain}

\section{Introduction}
The study of complex systems in physics has progressively utilized advanced mathematical models to clarify the underlying mechanisms of natural phenomena. This manuscript investigates two significant models: the Davey-Stewartson type system and the two-component third-order Burgers type systems. These models are closely linked to physical phenomena, providing valuable perspectives on a range of processes from fluid dynamics to wave propagation across different media.

The phenomenon of jams in congested traffic demonstrates a range of density waves, including distinct nonlinear wave patterns like solitons, triangular shocks, and kinks. These density waves can be modeled using nonlinear wave equations such as the Korteweg-de-Vries (KdV) equation, the Burgers equation, and the Modified KdV equation, as discussed in \cite{ph1bur}. Specifically, the Burgers equation is utilized to model the density wave within the stable region of traffic flow, employing nonlinear analysis techniques as elaborated in \cite{phy2bur}.

The two-dimensional generalization of the standard cubic one-dimensional\newline Schrödinger equation in water wave theory leads to what is known as the Davey-Stewartson systems, detailed in \cite{ph2DS}. The Davey-Stewartson equation, first documented in \cite{{orgnlDS}}, models how the amplitude of a surface wave interacts with the mean flow and is crucial in various physical contexts \cite{ph1DS}. Its ability to simulate wave interactions in two dimensions is vital for understanding the complex behaviors and instabilities of waves in such scenarios.

This research focuses on the integrability of such systems by calculating master symmetries. Since integrable systems are characterized by having an infinite number of symmetries, master symmetries are crucial for verifying their existence and confirming integrability, as noted in references \cite{olverbook, bookint,pfokas,pFuch, bookBlu}. We employ the $\mathscr{O}$-scheme, introduced by Wang \cite{Wang2015, sander, thJing}, to calculate these master symmetries. This method relies on the $\mathfrak{sl}(2,\CC)$ representation, initially discussed in \cite{pslfirst}. In essence, the $\mathscr{O}$-scheme indicates that for a homogeneous evolution equation $u_t=K[u]$, a master symmetry can be computed using the Lie bracket $\tau= \frac{1}{2\kappa}[x^2u_x+2\alpha xu, K]$, provided specific criteria are met.

In their paper \cite{talati2016two}, Talati and Turhan classified systems of the Burger type that feature a non-diagonal constant matrix for the leading-order term. The first three examples in Section 2 we examine in this work were introduced as new systems in their study. While they identified a master symmetry, our research reveals additional or a parameterized set of master symmetries, which facilitates the generation of triangular symmetries through specific linear combinations. On the other hand, the Davey-Stewartson (DS) equation was categorized in the works of \cite{huard2013classification} and \cite{DSc}. However, these studies did not discover the master symmetry that we introduce in our current work.
\newline
Some notation:
We consider an evolution differential equation of a dependent variable $u$ of the form $$u_t=K[u],$$
where $[u]$ means that the smooth function $k$ depends on $u$ and $x-$derivatives of $u$ up to a finite order.
An equation is termed ($n$+1)-dimensional if it involves $n$ spatial variables and a time variable $t$.

All smooth functions depending on $x, t, u,$ and $x-$derivatives of $u$ form a differential ring of differential functions $\curly{F}$ with the total $x-$derivation
$$D_x=\derf{}{x}+\sum_{k=0}^\infty u_{(k+1)x}\derf{}{u_{k}},$$
where $u_{k}=\partial_x^k u.$
For any two differential functions $P$ and $Q$, we define a bracket as
\begin{equation}\label{LieBracket}
[P,Q]=Q_*(P)-P_*(Q).
\end{equation}
where $P_*$ is the differential operator (Fr\'echet derivative) defined as
\begin{equation}\label{formula:FreDer}
P_*=\sum_j\derf{P}{u_{jx}}D^j_x.
\end{equation}
For example, If $K[u]=uu_{x}+u_{xx}$ then $K_*=u_x+uD_x+D_{x}^2$ is its Fr\'echet derivative. One can check, that $[P,Q]=Q_*(P)-P_*(Q)$ is a Lie bracket.

We call an evolution equation $\alpha-homogeneous$ if $[xu_x+\alpha u,K]=\kappa K$ for some constants $\alpha$ and $\kappa$. In the case of a homogeneous equation with a scaling $h=2(xu_x+\alpha u)$, the elements $e=u_x, f=-(x^2u_x+2\alpha xu)$ and $h$ form an $\mathfrak{sl}(2,\CC)$ algebra (Lemma 1 in \cite{Wang2015}).
\begin{definition}
A $t$-independent differential function $\tau$ is a master symmetry of $u_t=K$ if and only if
\begin{equation}
[[\tau,K],K]=0 \;\text{provided}\; [\tau,K] \neq 0.
\end{equation}
\end{definition}

Master symmetry is a differential function $\tau \in \curly{F}$ whose adjoint action $\ad_{\tau}=[\tau, \cdot]$ maps a symmetry to a new symmetry \cite{bookdorf}.
The following theorems provides a constructive approach to find a master symmetry and time-dependent symmetries. Their proof can be found in Theorem 2 and 3 of \cite{Wang2015}.
\begin{theorem} \label{th:ComputingMastersymm}
For a homogeneous evolution equation $u_t=K[u]$ satisfying
$$[u_x,K] = 0, \hspace*{.3cm} [xu_x + \alpha u,K] = \kappa K, \kappa > 1,$$
with a certain constant $\alpha$, let
\begin{equation}\label{formula:tau}
\tau= \frac{1}{2\kappa}[x^2u_x+2\alpha xu, K]
\end{equation}
and $a_{n+1} = [\tau, a_n]$ with $a_0 = e = u_x.$ If
\begin{itemize}
\item $[[\tau, K], K]=0$ and
\item there is a Lie subalgebra $\mathfrak{h}$ such that $a_n \in \mathfrak{h}$ for all $n = 0, 1, 2, \cdots $ and, moreover, for any $P,Q \in \mathfrak{h}$
satisfying $[P,e]=[Q,e]=[P,K]=[Q,K]=0$, it follows that $[P,Q] = 0$,
\end{itemize}
then $\tau$ is a master symmetry of the equation $u_t=K[u]$, i.e., all $a_n$ mutually commute.
\end{theorem}\label{th:tdepsymm}
\begin{theorem}
Let the homogeneous equation $u_t=K[u]$ satisfy the condition of the Theorem \ref{th:ComputingMastersymm}. Let the evolution vector field $w$ satisfy
\begin{equation*}
\ad_e w= f, \hspace*{0.6cm} \ad_K w= \frac{1}{2-4\kappa}\ad^2_f K .
\end{equation*}
Then
\begin{equation*}
\dfrac{\partial^j}{\partial t^j}\exp(-t \, \ad_K)(\ad^n_f\, w)
\end{equation*}
are time-dependent symmetries for $n=0,1, \cdots$ and $j=0,1,\cdots n+2$.
\end{theorem} 
We commence with a straightforward example to demonstrate the functioning of the theorem.
\begin{example}\label{example:BurgersEq}
Consider the Burgers' equation of order 3
\begin{equation}\label{eq:Brgr_eq_thrd_Ordr}
u_t=K=B_3[u]=u_{3}+3uu_{2}+3u^2u_x+3u_x^2.
\end{equation}
To determine whether this system is homogeneous, we need to perform the Lie bracket \eqref{LieBracket} with the scaling symmetry $h=xu_x+u$,
\begin{align*}
[h,u_t]&=u_{t*}(h)-h_*(u_t),\\
&=(6uu_x + (3u_2+3u^2 + 6u_x)D_x+3uD_2+D_3)(h)-(1+xD_x)(u_t),\\
&=(3xu^2u_2 + 6xuu_x^2 + 3xuu_3 + 9xu_xu_2 + 12u^2u_x\\
& \hspace*{.5in} + xu_4 + 12uu_2 + 12u_x^2 + 4u_3)-(3xu_x^2u_2 + 6xuu_x^2 \\
&\hspace*{.5in} + 3xuu_3 + 9xu_xu_2 + 3u^2u_x + xu_4 + 3uu_2 + 3u_x^2 + u_3),\\
&=9u^2u_x+9uu_{2}+9u_x^2+3u_{3}=3K.
\end{align*}
This shows that the system is homogeneous with $\alpha=1$. We created a function ``LieBracket" in Maple to compute such calculations which is mentioned in the appendix at the end. Using this function, we compute $\tau$ according to Theorem \ref{th:ComputingMastersymm} 
\begin{align*}
\tau &=\frac{1}{6} [x^2u_x+2xu, K],\\
&=3xu^2u_x+3xuu_{2}+3xu_x^2+u^3+xu_{3}+5uu_x+2u_{2}.
\end{align*}
To verify whether $\tau$ is a master symmetry, first we compute
\begin{align*}
[\tau,K]= -15u^4&u_x-30u^3u_{2}-90u^2u_x^2-30u^2u_{3}-150uu_xu_{2}\\
&-45u_x^3-15uu_{4}-45u_xu_{3}-30u_{2}^2-3u_{5},
\end{align*}
 and further calculation shows that $[[\tau,K],K]=0$, confirming that $\tau$ is a master symmetry.
\end{example}
\section{Two-component system}
Mikhailov, Shabat, and Yamilov in \cite{pshabYami} initiated the systematic classification of multi-component integrable equations. Talati and Tuhan provided a classification for $(1,1)$-homogeneous systems in \cite{talati2016two} that were previously unclassified with respect to the matrix of leading order terms in \cite{pfour, pTsu, psw}.
We will compute master symmetries for three systems in Examples \ref{system12}, \ref{system13} and \ref{system14}. These systems appears to be new among the list of  eight systems found by Talati and Turhan . Later in the section, we also consider two new systems in Examples \ref{systemNew13} and \ref{systemNew14} from the ongoing work of Wang (et al.) \cite{WangInprep}. The calculations were performed using Maple, with the relevant code available in the Appendix at the end.
First we extend the definition of the Lie bracket (\ref{LieBracket}) for two component systems systems.
\begin{definition}
For any two evolutionary vector fields with characteristics 
$\colvec{2}{P}{Q}$ and $\colvec{2}{J}{K}$, the Lie bracket is defined as
\[
\begin{bmatrix}
  \begin{pmatrix}
  P\\
  Q
\end{pmatrix}  & \begin{pmatrix}
  J\\
  K
\end{pmatrix}

\end{bmatrix}
=
\begin{pmatrix}
   J_{u}^*(P)+J_{v}^*(Q)-P_{u}^*(J)-P_{v}^*(K)\\
    K_{u}^*(P)+K_{v}^*(Q)-Q_{u}^*(J)-Q_{v}^*(K)
\end{pmatrix}.
\]
where $A_{u}^*(B)$ means the Fr\'echet derivative of $A$ with respect to $u$ acting on $B$.
\end{definition}

\begin{example}\label{system12}
This examples comes from a noval system which appears as System $(12)$ in \cite{talati2016two}, 
\begin{equation}
\left.
\begin{aligned} 
u_t&=B_3[u]+2\epsilon(v_{2}+3uv_x+2vu_{1}+2u^2v)_x\\
v_t&=v_{3}+3uv_{2}+6u_xv_x+3u^2v_x-\epsilon(4vv_{2}-v_x^2+8u_xv^2+8uvv_x+4u^2v^2)
\end{aligned}
\right\}=K.
\end{equation}
Here, we provide a comprehensive, step-by-step explanation of the process.
\begin{itemize}
\item Initially, we verify the homogeneity of the system.
\item Next, we calculate $\tau$ following the method outlined in Theorem \ref{th:ComputingMastersymm}. We then confirm whether $[[\tau, K], K] = 0$ to ascertain that $\tau$ is indeed a master symmetry.
\item Perform some analysis on the master symmetry.
\item Determine time-dependent symmetries utilizing the process described in Theorem \ref{th:ComputingMastersymm}.
\end{itemize}
\textbf{Verify homogeneity}\\
The system is $(1,1)$ homogeneous, meaning, the following Lie brackets gives
\begin{equation}
\begin{bmatrix}
  \begin{pmatrix}
  xu_x+u\\
  xv_x+v
\end{pmatrix}  & \begin{pmatrix}
  u_t\\
  v_t
\end{pmatrix}

\end{bmatrix}
=
3\begin{pmatrix}
   u_t\\
   v_t
\end{pmatrix},
\end{equation}
and the underlying Lie algebra is generated by the elements 
$$e=\colvec{2}{u_x}{v_x},\,h=2\colvec{2}{xu_x+u}{xv_x+v},\,f=-\colvec{2}{x^2u_x+2xu-1}{x^2v_x+2xv+\gamma}.$$
Notice that $f=\colvec{2}{f_1}{f_2}$ has a free parameter $\gamma$.\\
\textbf{Construction of $\tau$}\\
Now, we compute $\tau=\colvec{2}{\tau_1}{\tau_2}$ using Theorem \ref{th:ComputingMastersymm}
\begin{equation}\label{ex:tauk}
\begin{pmatrix}
   \tau_1\\
   \tau_2
\end{pmatrix}=
\frac{1}{6}\begin{bmatrix}
  \begin{pmatrix}
  f_1\\
  f_2
\end{pmatrix}  & \begin{pmatrix}
  u_t\\
  v_t
\end{pmatrix}
\end{bmatrix}.
\end{equation}
This results in a one-parameter family of differential functions with
\begin{align*}
\tau_1=-xu_t&-\tfrac{2}{3} \gamma\epsilon u_{2}-\tfrac{3}{2}u_{2}-3\epsilon v_{2}-\tfrac{4}{3}\gamma\epsilon uu_x-6\epsilon vu_x\\
&-4uu_x-8\epsilon uv_x-4\epsilon u^2v-u^3,\\
\tau_2=-xv_{t}&+\tfrac{2}{3}\gamma\epsilon v_{2}-\tfrac{3}{2}v_{2}+\tfrac{8}{3}\gamma\epsilon vu_x-2vu_x+\tfrac{4}{3}\gamma\epsilon uv_x\\
&+2\epsilon vv_x-4uv_x+\tfrac{4}{3}\gamma\epsilon u^2v+4\epsilon uv^2-u^2v.
\end{align*}
Once can check that $[[\tau,K],K]=0$, therefore $\tau$ is a master symmetry as required.\\
\textbf{Analysis of master symmetries}\\
Setting $\gamma=\tfrac{3}{4}$ results in the same master symmetry $\textbf{M}=\colvec{2}{M_1}{M_2}$ (up to sign) as in \cite{talati2016two} where,
\begin{align*}
M_1= xu_{3}&+2\epsilon xv_{3}+4\epsilon xvu_{2}+3xuu_{2}+2u_{2}+6\epsilon xuv_{2}+3\epsilon v_{2},\\
&+3xu_x^2+10\epsilon xu_xv_x+8\epsilon xuvu_x+3xu^2u_x+6\epsilon vu_x\\
&+5uu_x+4\epsilon xu^2v_x+8\epsilon uv_x+4\epsilon u^2v+u^3,\\
M_2=xv_{3}&-4\epsilon xvv_{2}+3xuv_{2}+v_{2}+6xu_xv_x-8\epsilon xv^2u_x\\
&+\epsilon xv_x^2-8\epsilon xuvv_x+3xu^2v_x-2\epsilon vv_x+3uv_x\\
&-4\epsilon xu^2v^2-4\epsilon uv^2.
\end{align*}
 Since, $\textbf{M}$ and $\tau$ are dependent master symmetries, we can construct a lower order master symmetry $\mathcal{M}=\textbf{M}+\tau$ with
\begin{align*}
\curly{M}_1=-\tfrac{2}{3}\gamma&\epsilon u_{2}+\tfrac{1}{2}u_{2}-\tfrac{4}{3}\gamma\epsilon uu_x+uu_x,\\
\curly{M}_2=-\tfrac{1}{2}v_{2}&+\tfrac{2}{3}\gamma\epsilon v_{2}-2vu_x+\tfrac{8}{3}\gamma\epsilon vu_x-uv_x+\tfrac{4}{3}\gamma\epsilon uv_x\\
&-u^2v+\tfrac{4}{3}\gamma\epsilon u^2v.
\end{align*}
Note that $\curly{M}_1$ relies solely on the variable $u$, and $\curly{M}_2$ is linear in $v$, indicating a triangular master symmetry. This symmetry can generate triangular symmetries through appropriate linear combinations of those derived from $\tau$ and $\curly{M}$. For example, the following linear combination 
\begin{equation}
S=\,\left( \frac{8}{3}\gamma\epsilon^2-12\epsilon\right)[\curly{M},K]-\left(\frac{8}{3}\gamma\epsilon^2-2\epsilon\right)[\tau,K],
\end{equation}
 generates a triangular symmetry $S=\colvec{2}{S_1}{S_2}$ with
\begin{align*}
S_1=\,8\epsilon^2&u_5-6\epsilon u_5+40\gamma\epsilon^2uu_4-30\epsilon uu_4+120\gamma\epsilon^2 u_3u_1-90\epsilon u_3u_1\\
&+80\gamma\epsilon^2u^2u_3-60\epsilon u^2u_3+80\gamma\epsilon^2 u_2^2-60\epsilon u_2^2+400\gamma\epsilon^2 uu_2u_1\\
&-300\epsilon uu_2u_1+80\gamma\epsilon^2u^3u_2-60\epsilon u^3u_2+120\gamma\epsilon^2u_1^3-90\epsilon u_1^3\\
&+240\gamma\epsilon^2 u^2u_1^2-180\epsilon u^2u_1^2+40\gamma\epsilon^2 u^4u_1-30\epsilon u^4u_1,\\
\\
S_2=\,8\gamma&\epsilon^2 v_5-6\epsilon v_5+40\gamma\epsilon^2uv_4-30\epsilon uv_4+80\gamma\epsilon^2u_3v_1\\
&-60\epsilon u_3v_1+120\gamma\epsilon^2v_3u_1-90\epsilon v_3u_1+80\gamma\epsilon^2 u^2v_3\\
&-60\epsilon u^2v_3+120\gamma\epsilon^2u_2v_2-90\epsilon u_2v_2+240\gamma\epsilon^2 uu_2v_1\\
&-180\epsilon uu_2v_1+320\gamma\epsilon^2uv_2u_1-240\epsilon uv_2u_1+80\gamma\epsilon^2 u^3v_2\\
&-60\epsilon u^3v_2+280\gamma\epsilon^2u_1^2v_1-210\epsilon u_1^2v_1+320\gamma\epsilon^2u^2u_1v_1\\
&-240\epsilon u^2u_1v_1+40\gamma\epsilon^2 u^4v_1-30\epsilon u^4v_1.
\end{align*}
This means that we can find a suitable transformation to linearise this system.\\

By the property of Master symmetry, the adjoint action of $\tau$ on $K$, i.e, $\ad^n_{\tau}K$ generates infinitely many symmetries, each of order $3+2n$. This can be proved by induction on $n$.\\
\textbf{Time-dependent symmetries}\\
For time-dependent symmetries we use Theorem \ref{th:ComputingMastersymm}. It requires us to find an element $w$ such that
\begin{equation*}
\ad_e w= f, \hspace*{0.6cm} \ad_K w= \frac{1}{2-4\kappa}\ad^2_f K .
\end{equation*}
We find this element by $w=-\cum f\,dx$ which provides
\[\colvec{2}{w_1}{w_2}=
\begin{pmatrix}
 \tfrac{1}{3} x^3u_x+x^2u-x  \\
 \tfrac{1}{3}x^3v_x+x^2v+\gamma x
\end{pmatrix}.
\]
This element satisfies the first condition $\ad_e w= f$. From Equation \eqref{ex:tauk}, we already know that $\tfrac{1}{6}\,[f, K]=\tau$, therefore the condition $\ad_K w= \frac{1}{2-4\kappa}\ad^2_f K $ takes the below form
\[
\begin{bmatrix}
\colvec{2}{u_t}{v_t}&
\colvec{2}{w_1}{w_2}  
\end{bmatrix}- \frac{-3}{5}
\begin{bmatrix}
  \begin{pmatrix}
  f_1\\
  f_2
\end{pmatrix}  & \begin{pmatrix}
  \tau_1\\
  \tau_2
\end{pmatrix}

\end{bmatrix}
=0 .\]
On inspection we find that for $\gamma=0$ or $\gamma=3$, i.e, both $w$ works
\[\colvec{2}{w_1}{w_2}=
\begin{pmatrix}
 \tfrac{1}{3} x^3u_x+x^2u-x  \\
 \tfrac{1}{3}x^3v_x+x^2v
\end{pmatrix}, \text {and} \colvec{2}{w_1}{w_2}=
\begin{pmatrix}
 \tfrac{1}{3} x^3u_x+x^2u-x  \\
 \tfrac{1}{3}x^3v_x+x^2v+3x
\end{pmatrix},
\]
Therefore,
\begin{equation*}
\dfrac{\partial^j}{\partial t^j}\exp(-t \, \ad_K)(\ad^n_f\, W)
\end{equation*}
are time-dependent symmetries for $n=0,1, \cdots$ and $j=0,1,\cdots n+2$.\\
\end{example}
In the following examples we continue in similar fashion.
\begin{example}\label{system13}
Consider the System
\begin{equation}
\left.
\begin{aligned} 
u_t=B_3[u]+\epsilon(v_{2}+2uv_x+vu_x+u^2v)_x,\\
v_t=v_{3}+6u_xv_x-\epsilon(vv_{2}-v_x^2+v^2u_x).
\end{aligned}
\right\}=K
\end{equation}
%
%
provided $[[\tau,K],K]=0$, i.e, we need to show $[\tau, K]$ is a symmetry of the system. We find that
\begin{align*}
\tau_1=-xu_{3}&-\epsilon xv_{3}-\epsilon xvu_{2}-3xuu_{2}-2u_{2}-2\epsilon xuv_{2}-\tfrac{4}{3}\epsilon v_{2}\\
&-3xu_x^2-3\epsilon xu_xv_x-2\epsilon xuvu_x-3xu^2u_x-\tfrac{4}{3}\epsilon vu_x\\
&-5uu_x-\epsilon xu^2v_x-\tfrac{7}{3}\epsilon uv_x-\epsilon u^2v-u^3,\\
\tau_2=-xv_{3}&+\epsilon xvv_{2}-v_{2}-6xu_xv_x+\epsilon xv^2u_x-\epsilon xv_x^2+\tfrac{1}{3}\epsilon vv_x\\
&-2uv_x+\tfrac{1}{3}\epsilon uv^2,
\end{align*}
and the maple computation (see appendix) indeed shows $[[\tau,K],K]=0$.

Master symmetry $\textbf{M}$ computed by Talati and Turhan \cite{talati2016two} is
\begin{align*}
M_1= xu_{3}&+\epsilon xv_{3}+\epsilon xvu_{2}+3xuu_{2}+2u_{2}+2\epsilon xuv_{2}+3\epsilon v_{2}\\
&+3xu_x^2+3\epsilon xu_xv_x+2\epsilon xuvu_x+3xu^2u_x+6\epsilon vu_x\\
&+5uu_x+\epsilon xu^2v_x+8\epsilon uv_x+4\epsilon u^2v+u^3,\\
M_2=xv_{3}&-\epsilon xvv_{2}+v_{2}+6xu_xv_x-\epsilon xv^2u_x\\
&+\epsilon xv_x^2-2\epsilon vv_x+3uv_x-4\epsilon uv^2.
\end{align*}
Similar to the previous example, here we can also find a lower order master symmetry $\curly{M}=\textbf{M}+\tau$ with order 2
\begin{align*}
\curly{M}_1=&\tfrac{5}{3}\epsilon v_{2}+\tfrac{14}{3}\epsilon vu_x+\tfrac{17}{3}\epsilon uv_x+3\epsilon u^2v,\\
\curly{M}_2=&uv_x-\tfrac{5}{3}\epsilon vv_x-\tfrac{11}{3}\epsilon uv^2.
\end{align*}

In this case, $\ad^n_\curly{M}$ generates symmetry of order $3+n$ whereas $\ad^n_\tau$ generates symmetry of order $3+2n$.
\end{example}
\begin{example}\label{system14}
The following system 
\begin{equation}
\left.
\begin{aligned} 
 u_t=B_3[u]+3\epsilon(v_{2}-vu_x)_x+3\epsilon^2(4vv_{2}+3v_x^2-v^2u_x)+12\epsilon^3v^2v_x,\\
v_t=v_{3}-3(uv_x-u^2v)_x+3\epsilon(3vv_{2}+3v_x^2-2(v^2u)_x+21\epsilon^2v^2v_x.
\end{aligned}
\right\}=K
\end{equation}
is $(1,1)$ homogeneous similar to our previous examples with the same $\kappa$ factor,
\[
\begin{bmatrix}
  \begin{pmatrix}
  xu_x+u\\
  xv_x+v
\end{pmatrix}  & \begin{pmatrix}
  u_t\\
  v_t
\end{pmatrix}

\end{bmatrix}
=
3\begin{pmatrix}
   u_t\\
   v_t
\end{pmatrix}.
\]
With $f=-\colvec{2}{x^2u_x+2xu+1}{x^2v_x+2xv}$, the following Lie bracket
\[
\frac{1}{6}\begin{bmatrix}
  \begin{pmatrix}
f_1\\
f_2
\end{pmatrix}  & \begin{pmatrix}
  u_t\\
  v_t
\end{pmatrix}

\end{bmatrix}
=
\begin{pmatrix}
   \tau_1\\
   \tau_2
\end{pmatrix},
\]
provides a master symmetry $\tau$ with
\begin{align*}
\tau_1=-xu_{3}&-3\epsilon xv_{3}+3\epsilon xvu_{2}-3xuu_{2}-\tfrac{5}{2}u_{2}-12\epsilon^2 xvv_{2}-6\epsilon v_{2}\\
&-3xu_x^2+3\epsilon xu_xv_x+3\epsilon^2 xv^2u_{1}-3xu^2u_x+4\epsilon vu_x-6uu_x-9\epsilon^2 xv_x^2\\
&-12\epsilon^3 xv^2v_x-18\epsilon^2 vv_x+\epsilon uv_x-4\epsilon^3v^3+\epsilon^2uv^2-u^3,\\
\tau_2=-xv_{3}&-9\epsilon xvv_{2}+3xuv_{2}-\tfrac{3}{2}v+3xu_xv_x+6\epsilon xv^2u_x-6xuvu_x\\
&-9\epsilon xv_x^2-21\epsilon^2 xv^2v_x+12\epsilon xuvv_x-3xu^2v_x-13\epsilon vv_x+3uv_x\\
&-7\epsilon^2 v^3+6\epsilon uv^2-3u^2v.
\end{align*}
After comparing the master symmetry \textbf{M} as in \cite{talati2016two},
\begin{align*}
M_1= xu_{3}&+3\epsilon xv_{3}-3\epsilon xvu_{2}+3xuu_{2}+2u_{2}\\
&+12\epsilon^2 xvv_{2}+3\epsilon v_{2}+3xu_x^2-3\epsilon xu_xv_x\\
&-3\epsilon^2 xv^2u_x+3xu^2u_x+6\epsilon vu_x+5uu_x,\\
&+9\epsilon^2 xv_x^2+12\epsilon^3 xv^2v_x+8\epsilon uv_x+4\epsilon u^2v+u^3\\
M_2=xv_{3}&+9\epsilon xvv_{2}-3xuv_{2}+v_{2}-3xu_xv_x\\
&-6\epsilon xv^2u_x+6xuvu_x+9\epsilon xv_x^2\\
&+21\epsilon^2 xv^2v_x-12\epsilon xuvv_x\\
&+3xu^2v_x-2\epsilon vv_x+3uv_x-4\epsilon uv^2,
\end{align*}
we find a lower order master symmetry $\curly{M}=\textbf{M}+\tau$ of order 2,
\begin{align*}
\curly{M}_1=&-\tfrac{1}{2}u_{2}-3\epsilon v_{2}-uu_x+10\epsilon vu_x+9\epsilon uv_x,\\
&-18\epsilon^2 vv_x+4\epsilon u^2v-4\epsilon^3 v^3+\epsilon^2 uv^2\\
\curly{M}_2=&-\tfrac{1}{2} v_{2}+6uv_x-15\epsilon vv_x-3u^2v\\
&+2\epsilon uv^2-7\epsilon^2 v^3.
\end{align*}
Here again we find that $\ad^n_\curly{M}$ generates symmetry of order $3+n$ whereas $\ad^n_\tau$ generates symmetry of order $3+2n$.

For time-dependent symmetries we find the element $w=-\cum f\,dx$ which provides
\[\colvec{2}{w_1}{w_2}=
\begin{pmatrix}
 \tfrac{1}{3} x^3u_x+x^2u+x  \\
 \tfrac{1}{3}x^3v_x+x^2v
\end{pmatrix}.
\]
Notice that the required constant of integrations are $0$ so that the following conditions are satisfied
\[\ad_e w= f,\quad
\begin{bmatrix}
\colvec{2}{u_t}{v_t}&
\colvec{2}{w_1}{w_2}  
\end{bmatrix}- \tfrac{-3}{5}
\begin{bmatrix}
  \begin{pmatrix}
  f_1\\
  f_2
\end{pmatrix}  & \begin{pmatrix}
  \tau_1\\
  \tau_2
\end{pmatrix}

\end{bmatrix}
=0.\]
Now,
\begin{equation*}
\dfrac{\partial^j}{\partial t^j}\exp(-t \, \ad_K)(\ad^n_f\, w)
\end{equation*}
are time-dependent symmetries for $n=0,1, \cdots$ and $j=0,1,\cdots n+2$.
\end{example}
So far all the two-component systems that we have considered have appeared in \cite{talati2016two}. The following two examples are from current work (in progress) of Wang et al. \cite{WangInprep}.
\begin{example}\label{systemNew13}
Consider the following system,
\begin{equation}
\overfullrule=0pt \left.
\begin{aligned}
u_t=&u_3+(\lambda-1)\left[2(u+v)v_2+6u^2v_1-12uvv_1-10v^2v_1+(u-v)^2(u+3v)(v+3u)\right]\\
 &+6uu_2+9u_1^2-6(vu_1)_x+(\lambda+2)\left[-v_1^2+4(u^2+v^2)u_1\right]+8(\lambda-4)uvu_1,\\
v_t=&\lambda v_3+(\lambda-1)\left[2(u+v)u_2+10u^2u_1+12uvu_1-6v^2u_1+(u-v)^2(u+3v)(v+3u)\right]\\
&+(2\lambda+1)\left[u_1^2+4(u^2+v^2)v_1\right]+3\lambda \left[2(uv_1)_x-2vv_2-3v_1^2\right]-8(4\lambda-1)uvv_1,\\
& \lambda \in \CC \setminus \{0,1\},
\end{aligned}
\right\}=K
\end{equation}
which is $(1,1)$ homogeneous with weight 3. We find a master symmetry $\tau$ by
\[\begin{pmatrix}
   \tau_1\\
   \tau_2
\end{pmatrix}=
-\frac{1}{6}\begin{bmatrix}
  \begin{pmatrix}
x^2u_1+2xu-\tfrac{1}{4}\\
x^2v_1+2xv+\tfrac{1}{4}
\end{pmatrix}  & \begin{pmatrix}
  u_t\\
  v_t
\end{pmatrix}

\end{bmatrix}
\]
with,
\begin{align*}
\tau_1=-xu_t&-\tfrac{3}{2}u_2+6vu_1-10uu_1+3uv_1+3vv_1-\lambda uv_1-\lambda vv_1+11u^2v\\
&-5uv^2-\lambda u^3+3\lambda v^3-3u^3-3v^3-3\lambda u^2v+\lambda uv^2,\\
\tau_2=-xv_t&-\tfrac{3}{2}\lambda v_2+vu_1+uu_1-3\lambda uu_1-3\lambda vu_1-6\lambda uv_1+10\lambda vv_1\\
&+u^2v-3uv^2-3\lambda u^3-3\lambda v^3+3u^3-v^3-5\lambda u^2v\\
&+11\lambda u v^2+4xu^3v-14xu^2v^2+4xuv^3.
\end{align*}
Since the order of the master symmetry is $3$, $\ad^n_{\tau}$ generates symmetry of order $3+2n$.
\end{example}
\begin{example}\label{systemNew14}
The following system 
\begin{equation}
\overfullrule=0pt \left.
\begin{aligned}
u_t=&u_3+3((u^2+v^2)u_1)_x-2(\lambda-1)uvv_2+(\lambda+2)uv_1^2+3(u^4+v^4)u_1\\
&+3uu_1^2-2(\lambda-1)(u^2+3v^2)uvv_1-2(2\lambda-5)u^2v^2u_1-(\lambda-1)uv^2(u^2+v^2)^2,\\
v_t=&\lambda(v_3)+2(\lambda-1)uvu_2+6\lambda uu_1v_1+9\lambda vv_1^2+3\lambda (u^2+v^2)v_2\\
&+(2\lambda+1)vu_1^2+2(\lambda-1)uv(v^2+3u^2)u_1+3\lambda(u^4+v^4)v_1+2(5\lambda-2)u^2v^2v_1\\
&+(\lambda-1)u^2v(u^2+v^2)^2.
\end{aligned}
\right\}=K
\end{equation}
is the only system so far which is $(\tfrac{1}{2},\tfrac{1}{2})$ homogeneous with weight 3.
%
Here, we find a master symmetry as
\[
-\frac{1}{6}\begin{bmatrix}
  \begin{pmatrix}
x^2u_1+xu\\
x^2v_1+xv
\end{pmatrix}  & \begin{pmatrix}
  u_t\\
  v_t
\end{pmatrix}

\end{bmatrix}
=
\begin{pmatrix}
   \tau_1\\
   \tau_2
\end{pmatrix},
\]
with
\begin{align*}
\tau_1=-xu_t&-\tfrac{3}{2}u_2-5u^2u_1-3v^2u_1-2u^3v^2-\tfrac{3}{2}uv^4-\tfrac{1}{2}u^5\\
&+\lambda uvv_1+\lambda u^3v^2+\lambda uv^4-3uvv_1,\\
\tau_2=-xv_t&-\tfrac{3}{2}\lambda v_2-\tfrac{3}{2}\lambda u^4v-2\lambda u^2v^3-3\lambda u^2v_1-5\lambda v^2v_1+uvu_1\\
&-\tfrac{1}{2}\lambda v^5+u^4v+u^2v^3-3\lambda uvu_1.
\end{align*}
which generates symmetries $S_n=\ad^n_{\tau}$ of order $3+2n$.
\end{example}
\section{Two-component (2+1)-dimensional partial differential equations}
\overfullrule=0pt The $\curly{O}$-scheme is a powerful tool for testing the integrability of $(1+1)$-dimensional nonlinear partial differential equations. We wish to extend this method for the $(2+1)$-dimensional case but we encounter few obstacles, chiefly non-locality, i.e, the appearance of the formal integral $D_x^{-1}$ or $D_y^{-1}$. The higher order symmetries and the equation themselves are non-local in their evolutionary form for integrable equations. This was noted by Mikhailov and Yamilov in \cite{mikhailov1998towards}, where they introduced a concept of \textit{quasi-local polynomials} to characterize nonlocalities.
The appearance of such operators forces us to extend the differential algebra $\curly{A}$. The naive approach is to adjoin all possible integrals, i.e, to construct the differential algebra $\curly{A}(D_x^{-1}, D_y^{-1})$. But now, any $f \in \curly{A}(D_x^{-1}, D_y^{-1})$ is a total derivative $f \in D\curly{A}(D_x^{-1}, D_y^{-1})$ and consequently,
\begin{equation*}
\curly{A}(D_x^{-1}, D_y^{-1})/ D\curly{A}(D_x^{-1}, D_y^{-1})=\CC.
\end{equation*}
This implies that all conservation laws are trivial. Therefore, such a construction seems not to be very fruitful.

Mikhailov and Yamilov also made another very important observation that the operators $D_x^{-1}$ and $D_y^{-1}$ never appear alone, but always in pairs like $D_x^{-1}D_y$ and $D_y^{-1}D_x$ for all known integrable equations and their hierarchies of symmetries. Based on this observation they introduced operators
\begin{equation}
\theta = D_x^{-1}D_y \hspace{.2cm} \text{and} \hspace{.2cm} \theta^{-1}=D_y^{-1}D_x,
\end{equation}
then many classes of equations and their symmetry hierarchies can be written without $D_x^{-1}$ and $D_y^{-1}$. In the following section we construct not only such an extension but also the  base ring $\curly{A}$ formally.
Now we are ready to extend our theory for the $(2+1)$-dimensional case. The below example can be found in the Huard  and Novikov \cite{huard2013classification}.
\begin{example}\label{ex:system3.5}
We follow the same routine procedure as before based on Theorem \ref{th:ComputingMastersymm}.\\
The following system,
\begin{equation}\label{ex:2plus1}
\left.
\begin{aligned} 
u_t=u_x\theta^{-1}u +\epsilon u_{2}+ (uv)_x,\\
v_t=(v\theta^{-1}u)_x-\epsilon v_{2}+vv_x.
\end{aligned}
\right\}=K
\end{equation}
is homogeneous with
\begin{equation}
\begin{bmatrix}
 \colvec{2}{xu_x}{xv_x+v} & \colvec{2}{u_t}{v_t}\end{bmatrix}=2\colvec{2}{u_t}{v_t}
.
\end{equation}
We obtain a master symmetry by the following action
\begin{align*}
\colvec{2}{\tau_1}{\tau_2}=&
\begin{bmatrix}
  \colvec{2}{x^2u_x}{x^2v_x+2xv}  & \colvec{2}{u_t}{v_t}
\end{bmatrix},\\
=\,&
4\begin{pmatrix}
   xu_t+\tfrac{1}{2}\epsilon u_x+\tfrac{1}{2}uv\\
    xv_t + v\theta^{-1}u-\tfrac{3}{2}\epsilon v_x +\tfrac{1}{2}v^2
\end{pmatrix}.
\end{align*}

Once we have a master symmetry $\tau$, we can proceed to compute a symmetry by computing the action of $\tau$ on the $K$, which yields
\begin{equation}
S=\begin{pmatrix}
   2 \epsilon^2u_3+3\epsilon (u_1v_1+u_2v+u_2\theta^{-1}u+u_1\theta^{-1} u_1)+3(uv)_x \theta^{-1}u\\
    +\tfrac{3}{2}uv\theta^{-1} u_1 +\tfrac{3}{2}u_1\theta^{-1}(uv)+3uvv_1+\tfrac{3}{2}u_1v^2+\tfrac{3}{2}u_1(\theta^{-1}u)^2\\
     2 \epsilon^2v_3-3\epsilon (u_2v+v_1^2+v_2\theta^{-1}u+v_1\theta^{-1} u_1)\\
     3vv_1\theta^{-1}u+\tfrac{3}{2}v^2\theta^{-1}u_1+\tfrac{3}{2}v\theta^{-1}(uv)_x+\tfrac{3}{2}v_1\theta^{-1}(uv)\\
     +\tfrac{3}{2}v_1v^2+3v\theta^{-1}u\theta^{-1}u_1+\tfrac{3}{2}v_1(\theta^{-1}u)^2
\end{pmatrix}.
\end{equation}
It is indeed a symmetry since one can compute that $[\tau,[\tau, K]=0$.

\end{example}
\section{Conclusions}
We introduce new findings—master symmetries for three novel two-component Burgers' type $(1+1)$-dimensional systems referenced in \cite{talati2016two} (Examples \ref{system12}, \ref{system13}, \ref{system14}), along with two additional systems from the recent research by Wang et al. \cite{WangInprep} (Examples \ref{systemNew13}, \ref{systemNew14}). In Section 3, we explore quasilocal polynomials, establishing an algebraic approach for examining $(2+1)$-dimensional systems, exemplified by a Davey-Stewartson type system (Example \ref{ex:system3.5}). We discovered that the master symmetry for an evolution equation $u_t = K[u]$ is expressed as $\tau = xu_t + r$, with some ``remaining'' differential function $r\in \curly{A}$. This observation suggests that investigating the master symmetry structure of an integrable equation, which explicitly includes the spatial variable $x$, could be a compelling direction for future research.
%

Bi-Hamiltonian systems cover a high percentage of the known integrable systems but there are integrable equations like Burgers' equation and Ibragimov-Shabat equation which fall outside of this category. Our approach can be used to construct time-dependent symmetries for these equations. These symmetries can be seen as a part of $\Sl(2,\CC)$-module, however our understanding of their appearance in the construction of symmetries is still limited. Constructing a scheme based on algebra of higher rank is  a promising direction of research which would allow us to study a wider class of PDEs. In future, extending this approach to integrable differential-difference and discrete systems could also be very important. 
\section{Acknowledgement}
The author acknowledges and appreciates the contributions of Professor J. P. Wang, the supervisor, for providing essential guidance during the preparation of this paper. Special thanks are extended to Professor Andy Hone for invaluable advice. This work is based on the results of the author's Ph.D. program at the School of Mathematics, Statistics, and Actuarial Science (SMSAS), University of Kent, UK and the support received from the School is gratefully acknowledged.

\bibliographystyle{amsplain}

\newpage
\section*{Appendix}\label{apndx:LieBrackets}
This section outlines the Maple code utilized for calculating master symmetries and time-dependent symmetries in the provided examples.
\begin{verbatim}
restart:
DD:=proc(f,n)#function to compute total derivative of f of order n
local w,ii,i,vv,ll;
w:=f;
for ii to n do
vv:=0;
ll:=sort(convert(map(op,indets(w)),list)):
 for i in ll do #this loop differentiate wrt u[i]
 vv:=vv+u[i+1]*diff(w,u[i]);
 od;
w:=diff(w,x)+vv; #this performs differentiation with respect to x
od;
RETURN(sort(expand(w)));
end:
LieBracket:=proc(P,Q)
# computing Lie bracket of the given functions P and Q
local n1,n2,i, A, B;
n1:=max(convert(map(op,indets(P,name)),set) minus indets(P,name));
#highest order derivative appearing in P
n2:=max(convert(map(op,indets(Q,name)),set) minus indets(Q,name));
A:=0; B:=0;
for i from 0 to n2 do
A:=A+diff(Q,u[i])*DD(P,i);#Frechet derivative of acting on P ie P_*(Q)
od;
for i from 0 to n1 do
B:=B+diff(P,u[i])*DD(Q,i);
od;
return(expand(A-B));#[P,Q]=D[Q](P)-D[P](Q)
end:
LieBracketSys:=proc(P,Q,R,T)
# computing Lie bracket for a system (a11=P, a12=R, a21=Q, a22=T)
local n1,n2,i, A, B;
n1:=max(convert(map(op,indets(P,name)),set) minus indets(P,name));
#highest order derivative appearing in P
n2:=max(convert(map(op,indets(Q,name)),set) minus indets(Q,name));
A:=0; B:=0;
for i from 0 to n2 do
A:=A+diff(Q,u[i])*DD(P,i);#Frechet derivative of acting on P ie P_*(Q)
od;
for i from 0 to n1 do
B:=B+diff(P,u[i])*DD(Q,i);
od;
return(expand(A-B));#[P,Q]=D[Q](P)-D[P](Q)
end:
\end{verbatim}
We use the below code for generalised two-component systems.
\begin{verbatim}
restart:
DD:=proc(f,n)#function to compute total derivative of f of order n 
local w,ii,i,vv,ll;
w:=f;
for ii to n do
vv:=0;
ll:=sort(convert(map(op,indets(w)),list)):
 for i in ll do #this loop differentiate wrt u[i]
 vv:=vv+u[i+1]*diff(w,u[i])+v[i+1]*diff(w,v[i]);
 od;
w:=diff(w,x)+vv; #this just do extra diff wrt x
od;
RETURN(sort(expand(w)));
end:
LieBracketFre:=proc(P,Q,R,T)# computing Lie bracket for 
a system (a11=P, a12=R, a21=Q, a22=T)
local hdu,hdv,i, A,L,K, B,hdU,hdV,j, C, D,M,N;

L:=convert(map(op,subs(seq(u[i]=0, i = 0 .. 10),indets(P,name))),set)
minus subs(seq(u[i]=0, i = 0 .. 10),indets(P,name));
L:={op(L),0};
hdv:=max(L);#highest order derivative of v appearing in P
K:=convert(map(op,subs(seq(v[i]=0, i = 0 .. 10),indets(P,name))),set)
minus subs(seq(v[i]=0, i = 0 .. 10),indets(P,name));
K:={op(K),0};
hdu:=max(K);#highest order derivative of u appearing in P
A:=0; B:=0;
for i from 0 to hdu do
A:=A+diff(P,u[i])*DD(R,i);#Frechet derivative of P wrt u acting on R
od;
for i from 0 to hdv do
B:=B+diff(P,v[i])*DD(T,i);#Frechet derivative of P wrt v acting on T
od;
return(simplify(A+B));

#[P,Q]=D[Q](P)-D[P](Q)
end:
LB:=proc(P,Q,R,T)
return(<LieBracketFre(P,Q,R,T),LieBracketFre(Q,P,R,T) >);#<P,Q>_*
end:
LieBracketSys:=proc(P,Q,R,T)
return(-LB(P,Q,R,T)+LB(R,T,P,Q));
end:
collectsort:=proc(f)#sorting with order u[i+1]>u[i] and u[i]=v[i]
local g,In,h;
h:=expand(f);
In:=seq([u[i],v[i]][],i=20..1,-1);
g:=sort(h, order=plex(In));
return(g);
end:
adjointP:=proc(n,p,q,r,t)#to compute adnP i.e n times
 adjoint of P=<p,q> first two components 
local M,L,i,j;
M:=LieBracketSys(p,q,r,t);
L:=convert(M,list);
for i from 1 to n-1 do
j:=nops(L);
L:=[op(L),op(convert(LieBracketSys(p,q,L[j-1],L[j]),list))];#recursion
od;
return(<collectsort(L[-2]),collectsort(L[-1])>);
end:

\end{verbatim}
\end{document}